%% file: main.tex
\documentclass[reprint,
superscriptaddress,
nofootinbib,
 amsmath,amssymb,
]{revtex4-2}

\usepackage{float}
\usepackage{graphicx}% Include figure files
\usepackage{lipsum}% http://ctan.org/pkg/lipsum
\usepackage{dcolumn}% Align table columns on decimal point
\usepackage{bm}% bold math
\usepackage{times}
\usepackage{multirow}
\usepackage{epstopdf}
\usepackage{color}
\usepackage[normalem]{ulem}
\usepackage{lineno}
\date{June 2020}

\usepackage{natbib}
\usepackage{graphicx}
\usepackage[final, inline, author=,nomargin]{fixme}
\fxsetup{theme=color}
\FXRegisterAuthor{pts}{anb}{\color{red}Pranava notes}      % creates \ptsnote{} command
\FXRegisterAuthor{content}{anb1}{\color{blue} Takeaway}      % creates \contentnote{} command

\newcommand{\nuebar}{\ensuremath{\overline{\nu}_e}}
\newcommand{\msq}{\ensuremath{\Delta m^{2}_{41}}\hspace{1pt}}
\newcommand{\sinsq}{\ensuremath{\sin^{2}(2\theta_{14})}\hspace{1pt}}

\begin{document}
%\preprint{APS/123-QED}
\title{Note on arXiv:2005.05301, `Preparation of the Neutrino-4 experiment on search for sterile neutrino and the obtained results of measurements'}
\input{Authors.tex}

\begin{abstract}
\normalsize
We comment on the claimed observation of sterile neutrino oscillations by the Neutrino-4 collaboration.
Such a claim, which requires the existence of a new fundamental particle, demands a level of rigor commensurate with its impact. 
The burden lies with the Neutrino-4 collaboration to provide the information necessary to prove the validity of their claim to the community.
In this note, we describe aspects of both the data and analysis method that might lead to an oscillation signature arising from a null experiment and describe additional information needed from the Neutrino-4 collaboration to support the oscillation claim.
Additionally, as opposed to the assertion made by the Neutrino-4 collaboration, we also show that the method of `coherent summation' using the $L/E$ parameter produces similar results to the methods used by the PROSPECT and the STEREO collaborations. 
\end{abstract}
\begin{titlepage}
\maketitle
\end{titlepage}
\section{Neutrino-4 Experiment}
\contentnote{Give a brief intro to N4 experiment.\\}
%\section{Neutrino-4 and the Search for Short-baseline Neutrino Oscillations}
Neutrino-4 is a reactor neutrino experiment designed to search for short-baseline sterile neutrino oscillations motivated primarily by the Reactor Antineutrino Anomaly~\cite{Huber11,Mention:2011rk}.
The 1.8\,m$^3$ Gd-doped liquid scintillator detector is divided into 50 sections consisting of 10 rows and 5 sections per row each of size 0.225\,m~$\times$ 0.225\,~m$\times$ 0.85\,m. 
With the first and last rows of the detector used as active veto, the detector is composed of a 1.42 m$^3$ active volume.
The detector is mounted on a movable platform enabling a baseline coverage of 6\,m to 12\,m from the reactor.

\subsection*{Oscillation Search Strategy}
\contentnote{N4 oscillation search strategy in brief. The equations used in this section are referred later in the paper.}
The strategy to search for sterile neutrino oscillations described in Ref.~\cite{Serebrov:2020May,Serebrov:2020,Serebrov:2020rhy} is presented here for the sake of completeness. 
For a given baseline~($L$) and energy~($E$), the theoretical and experimental rates are defined as 
\begin{equation}
    R_{i,k}^{\text{th}}=\frac{1-\sinsq \cdot \sin^2(1.27 \msq \cdot L_{k}/E_{i})}{K^{-1}\sum_{k}^{K}[1-\sinsq \cdot \sin^2(1.27 \msq \cdot L_{k}/E_{i})]} \\
    \label{eq:Rth}
\end{equation}
\begin{equation}
    R_{i,k}^{\text{exp}}=\frac{N(E_{i},L_{k}) \cdot L^2_{k}}{K^{-1}\sum_{k}^{K}N(E_{i},L_{k})\cdot L^2_{k}}
    \label{eq:Rexp}
\end{equation}
where $N$ is the measured rate, $i \text{ and } k$ are the energy and baseline indices respectively, and~\sinsq~and~\msq~are the oscillation parameters. 
A binned test statistic is then defined as 
\begin{equation}
    \chi^2(\sinsq,\msq)=\sum_{i,k}[(R^{\text{exp}}_{i,k}-R^{\text{th}}_{i,k})^2/(\Delta R^{\text{exp}}_{i,k})^2]
        \label{eq:chi2}
\end{equation}
where the energy and baselines are collapsed into $L/E, N(L/E_j)=N(E_i,L_k)$ and is described as `coherent summation'.

Neutrino-4 started data taking in June 2016 and has been collecting data since.
%An analysis performed on the data collected for 2 years which included 480 days of reactor on data and 278 days of reactor off indicated short baseline oscillations with the parameters (~\msq,~\sinsq)$\approx$(7.3 eV$^2$,0.39) at 2.8$\sigma$ C.L.
%The results were published in the JETP Letter~\cite{Serebrov:2018vdw} in 2019. 
The latest posting on the arXiv preprint~\cite{Serebrov:2020} repository includes analysis performed on 720 reactor-on days and 417 reactor-off days.
The data was divided into three different groupings all consisting of 24 baseline bins but varying in energy bin widths of 125, 250, and 500 keV. 
These bins were then collapsed into \textit{L/E} bins by merging\footnotetext{The bin widths and procedure of merging bins into groups is unclear from the article~\cite{Serebrov:2020May}.} adjacent points into groups of 8, 16 and 32 respectively. 
The final fit using Eq.~\ref{eq:chi2} was performed on an average of the three cases.
The corresponding \textit{L/E} distribution and the best fit oscillation parameters are shown in Figure 47 of Ref.~\cite{Serebrov:2020May}.
%Using this binning and the test statistic defined in Eq.~\ref{eq:chi2}, it is claimed that the data from Neutrino-4 favors short baseline oscillations corresponding to the oscillation parameters~(\msq,\sinsq)$\approx$(7.26 eV$^2$,0.38) at 3.5$\sigma$ C.L.
%Note that a more recent arXiv posting~\cite{Serebrov:2020rhy} with two of the Neutrino-4 collaborators as authors have updated the best fit oscillation parameters to $\approx$(7.25 eV$^2$,0.26) and the confidence level to 3.0$\sigma$. 
%The same data used in Ref.~\cite{Serebrov:2020} was binned in energy intervals of widths 125, 250, and 500 keV while keeping baseline bins identical. 

\section{Use of Proper Statistical Method for Oscillation Search}
\contentnote{Demonstrate clearly that no matter how good the data are, the significance of oscillation is overestimated if a wrong statistical approach is used. In other words show that even if we completely trust the data, an oscillation signature can't be taken seriously unless the right approach is used. This is because if the detector resolution(and binning) $\sim \msq$, any statistical fluctuations in data looks like a deviation caused by the oscillations of one of the infinite possible frequencies.}
The importance of using the right statistical approach in search for oscillation signatures has been widely discussed in literature~\cite{FC,Gaussian_CLS}.
In particular we call the reader's attention to Ref.~\cite{Agostini:2019jup} which among other things discusses the possibility of over-estimation of the significance of results when using the incorrect test statistic by short baseline disappearance experiments.
Sterile neutrinos at \msq $ \sim \text{eV}^2$ would induce oscillations of high frequencies that could generate variations of the order or narrower than resolution of a typical short baseline reactor antineutrino detector.
A consequence of high frequency oscillations is that the fits performed using a binned test statistic and finite data size prefers the oscillation hypothesis over the no oscillation hypothesis. 
This is the case even when the data does not result from true oscillations. 
The reason for this is that within the given wide range of oscillation parameter space, there always exist oscillations of some frequency that fits well to the statistical fluctuations in the data. 
Additionally, as described in detail in Ref.~~\cite{FC}, the regions in parameter space close to the bounded regions in \sinsq have incorrect coverage if the confidence intervals are assigned using Wilks' theorem.
Consequently, the test statistic distribution described in Eq.~\ref{eq:chi2} will not follow the standard $\chi^2$ distribution with two degrees of freedom under the Wilks' theorem assumption.

\begin{figure}
\includegraphics[width=0.48\textwidth]{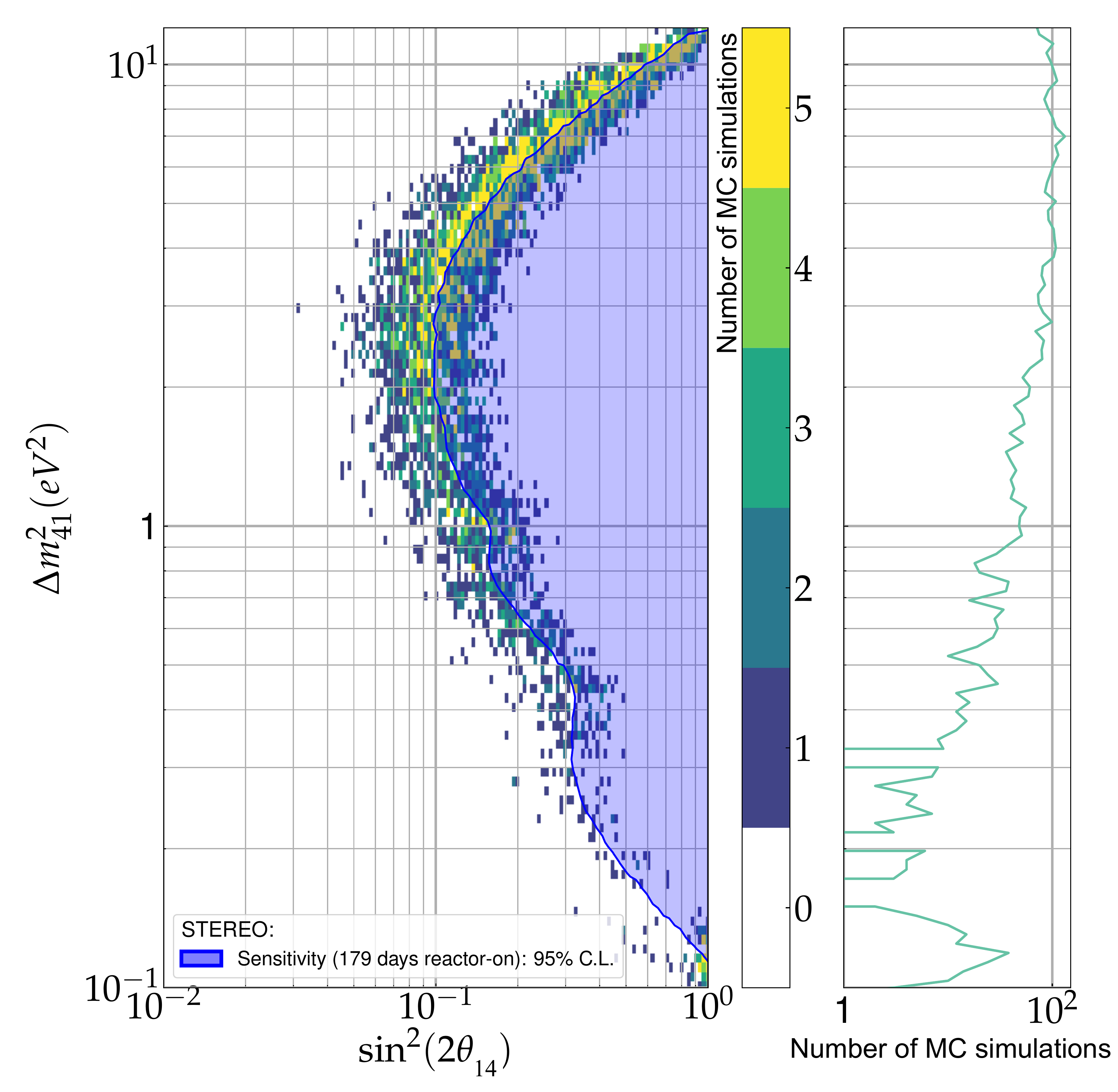}
\caption{The best-fit oscillation parameters (\sinsq,\msq) for simulated Monte Carlo experiments generated using the duration, detector response, and uncertainties based on the 179-day dataset published by the STEREO experiment~\cite{STEREO2019}. Even though the simulated MC experiments are generated under the assumption of no sterile neutrino oscillations, the best-fit oscillation amplitude \sinsq is never zero. In particular, the plot on the right shows the marginalized best-fit points in \msq  demonstrating a preference to oscillations at high frequencies.}
\label{fig:overfit}
\end{figure}

\begin{figure}
\includegraphics[width=0.48\textwidth]{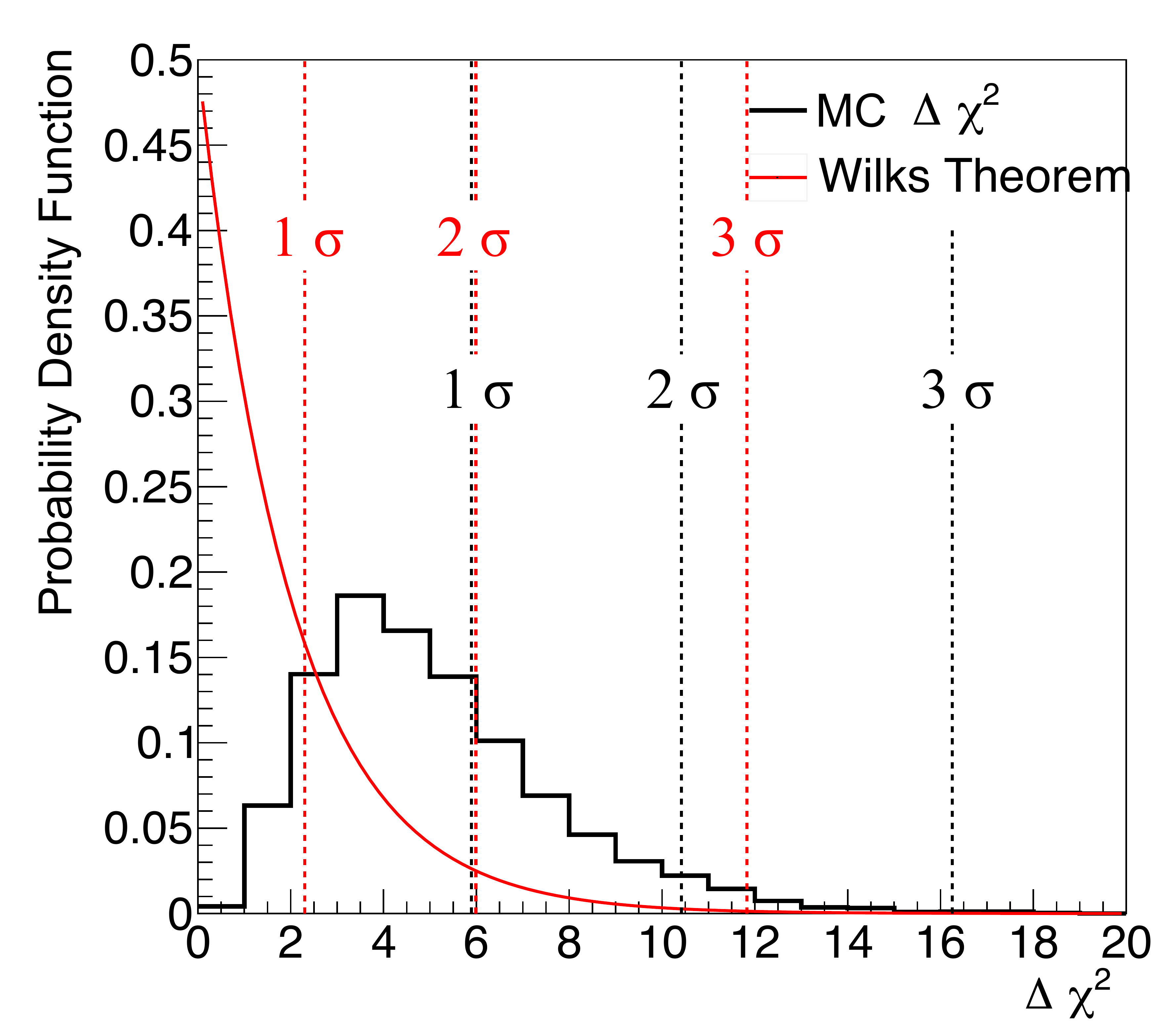}
\caption{MC generated $\Delta \chi^2$ distribution compared with the  Wilks' theorem-predicted values for no oscillation hypothesis. An undercoverage by the Wilks' theorem  values for the simulated MC datasets shown in Fig.~\ref{fig:overfit} for all values of $\Delta \chi^2$. Also shown in the figure in red and black vertical dashed lines are 1,2, and 3 $\sigma$ values for the standard $\chi^2$ and data-generated $\Delta \chi^2$ respectively emphasizing the key point that assigning significance based on standard $\chi^2$ leads to significantly skewed results.}
\label{fig:Deltachi2}
\end{figure}
The above situation is illustrated using simulated Monte Carlo~(MC) datasets generated assuming 179~(235) reactor-on~(-off) days of STEREO experiment and all the relevant statistical and systematic uncertainties.
This corresponds to the same size of dataset used by STEREO's recent~\cite{STEREO2019} sterile neutrino search.
%In a regime where the detectors search for oscillations in a constrained oscillation parameter space~(\sinsq$\geq 0$, \msq$\geq 0$) and at oscillation frequencies of the order of their energy and position resolutions, an over-fit to the fluctuations in data could lead to an oscillation-like signature.
Fig.~\ref{fig:overfit} shows the best-fit oscillation parameters for fits performed on 5000 null oscillated simulated MC experiments.
All best-fit oscillation parameters show a non-zero value of \sinsq with a preference to high frequencies where the oscillations are in the range comparable to the bin width of the analysis.
For the same simulated MC datasets, the $\Delta \chi^2$ distribution is shown in the Fig.~\ref{fig:Deltachi2} where $\Delta \chi^2$ is defined as the difference in $\chi^2$ generated using the best-fit and the true~(null oscillated) parameters.
It shows that this distribution is significantly different from the Wilks' theorem predicted $\chi^2$ distribution for two degrees of freedom. 
It is worthwhile to note that:
\begin{itemize}
    \item The median $\Delta \chi^2$ of simulated MC distribution is at 4.64 which corresponds to a 90 \% confidence level~(C.L.) under Wilks' theorem assumption. In other words, in absence of oscillations, data interpreted through Wilks' theorem would disfavor the null oscillation hypothesis at 90 \% C.L. half of the time.
    \item The $\Delta \chi^2$ of 11.83 corresponding to Wilks' theorem 3$\sigma$ includes only 98\% of the simulated MC experiments. This implies that even in absence of sterile neutrino induced oscillations, 1 in every 50 experiments would wrongly disfavor the null oscillation hypothesis at 3$\sigma$. 
\end{itemize}

This shows that experiments are predisposed to assign high significance to null oscillation exclusion if relying on standard $\chi^2$ distribution and further shows that a proper statistical approach is a prerequisite for any discovery claim.
Considering the importance of assigning proper significance, both PROSPECT and STEREO approach this issue by assigning the significance to the measured result by comparing the data-suggested $\Delta \chi^2$ to the $\Delta \chi^2$ distribution generated using experiment-specific simulated MC datasets. 
We likewise suggest that any experiment searching for short baseline sterile neutrinos --- especially an experiment claiming oscillation observation --- to use a similar MC-based approach so as not to overestimate the significance of their result. 
We were not able to find the use of MC-based statistical approach in any of the Neutrino-4 published results~\cite{Serebrov:2018vdw,Serebrov:2020,Serebrov:2020May,Serebrov:2020rhy}.
Thus, we encourage Neutrino-4 experiment to demonstrate the oscillation claim using a proper statistical approach for their analysis.

\section{Systematic Effects Mimicking Oscillation}
\contentnote{The previous section clearly shows that the significance of oscillations could be inflated if simulated MC datasets are not used. When coupled with systematic affects that could mimic oscillations, that would lead to oscillation signature.} 
In the previous section, we demonstrated that statistical fluctuations could be mistaken for an oscillation signature in absence of a proper statistical approach.
The problem could be exacerbated if there are unidentified, and thus unaccounted for, oscillation-mimicking systematic effects. 
This can be especially true for an experiment with a small detector located in close proximity to a reactor with little overburden.
Here we point out two such key systematic effects which are not discussed by the Neutrino-4 experiment.  
%It doesn't appear that any systematic uncertainties are included in defining the test statistic in Eq.~\ref{eq:chi2}. 
%Here we address some of the key systematic effects that could impact an experiment like Neutrino-4.
%several issues associated with Neutrino-4 data and the analysis techniques utilized. 
%\subsection*{Response of Detector to \nuebar}
%If the exposure at each baseline, the baseline- and energy-dependent detector response of the detector are accounted properly, the
%A few observations could be made about the Eqs.~\ref{eq:Rth} and~\ref{eq:Rexp}.  
%Consider the definitions of  and the test statistic in Eq.~\ref{eq:chi2}.

\contentnote{There are complex correlations between the energy spectrum, position, and efficiency.}
Neutrino-4 uses gadolinium-doped liquid scintillator~(LS) as the target and inverse beta decay~(IBD) mechanism for detecting neutrinos. 
The energy of the positron produced in the IBD interaction acts as a proxy for the \nuebar~energy while the neutron produced in the interaction captures on Gadolinium producing 3 to 4 Mev gamma rays and is used to establish a coincident signal.
In a segmented detector of scale $\sim$1 meter like Neutrino-4, the IBD positron and/or the annihilated $\gamma$s lose some of their energy through escape or energy deposition in inactive volume. 
This leads to an energy spectrum that is position-dependent since the IBD interactions taking place close to the edge of the detector have a higher fraction of escape energy.
Additionally, since the attenuation length of high energy $\gamma$s produced from neutron capture on Gadolinium is $\sim$15 cm, the neutron capture efficiency also varies within the detector volume.
These edge effects could induce complex correlations between energy and efficiency which could induce an oscillation-like signature.
The situation is further complicated by the fact that the detector is mobile and each detector segment spans over multiple baselines.    
However, these complex detector effects can be accounted for in the oscillation search by using a fully validated detector MC simulation. 
While it is not clear that these concerns introduce false oscillations, they will certainly effect the estimation of signal significance, and must be accounted for in a full analysis.
There is no indication that Neutrino-4 has incorporated their detector MC simulation in the theoretical rates from the Eq.~\ref{eq:Rth}. 
Therefore, we request Neutrino-4 experiment to provide more details on the calculation of their theoretical rates alongside a detector MC simulation that is fully validated using calibration data.

\contentnote{Point out that the reactor facilities have complicated backgrounds. Neutrino-4 backgrounds already look like oscillations. Not surprising if the signal also sees oscillations.}
Short baseline reactor neutrino experiments have to overcome challenging --- often position- and energy-dependent --- background environments in the search for \nuebar~oscillations. 
%It is further complicated by the fact that both reactor and cosmogenic backgrounds are position- and detector location-dependent.
The correlated cosmogenic backgrounds can be measured during the reactor-off period and can be scaled to reactor-on period and subtracted from the reactor-on data.
In principle, this works well if the detector has a good signal-to-background~(S:B) ratio, the reactor-off duration accounts for a significant amount~($\sim$50 \%) of data-taking, and there exists no variations~\footnote{There will still be atmospheric-dependent background fluctuations which can be estimated~\cite{PROSPECT_First_Search,STEREO2019,PhysRevD.93.112006} based on the correlations between atmospheric conditions and the IBD-like backgrounds during the reactor-off period.} in cosmogenic backgrounds between reactor-on and -off periods.
As discussed on page 8 in Ref.~\cite{Serebrov:2020}, in an earlier analysis, a fit performed using only 278 days of reactor-off data was found to yield an oscillation parameter at 99\% C.L~($\sim$2.6$\sigma$ C.L).
This could be an artifact of assigning incorrect significance based on using standard $\chi^2$ as discussed in the previous section.
Conversely, in conjunction with the low S:B ratio of $\sim0.5$ of Neutrino-4 experiment~\cite{2019tcg}, it could also suggest the possibility that the oscillation signature indicated by fluctuations in the cosmogenic backgrounds gets enhanced with addition of reactor-on data.
It is also possible that this is unrelated to the oscillation signature suggested by the IBD data.  
To disambiguate between various scenarios, we encourage the Neutrino-4 collaboration to provide more details on backgrounds and the background subtraction procedure employed in calculating the experimental IBD spectrum shown in Eq.~\ref{eq:Rexp}.

\section{Oscillation Signature Using L/E Oscillations}
\contentnote{Show that the oscillation search done using L/E yields similar results to the one performed using LvsE bins.}
\begin{figure}
\includegraphics[width=0.5\textwidth]{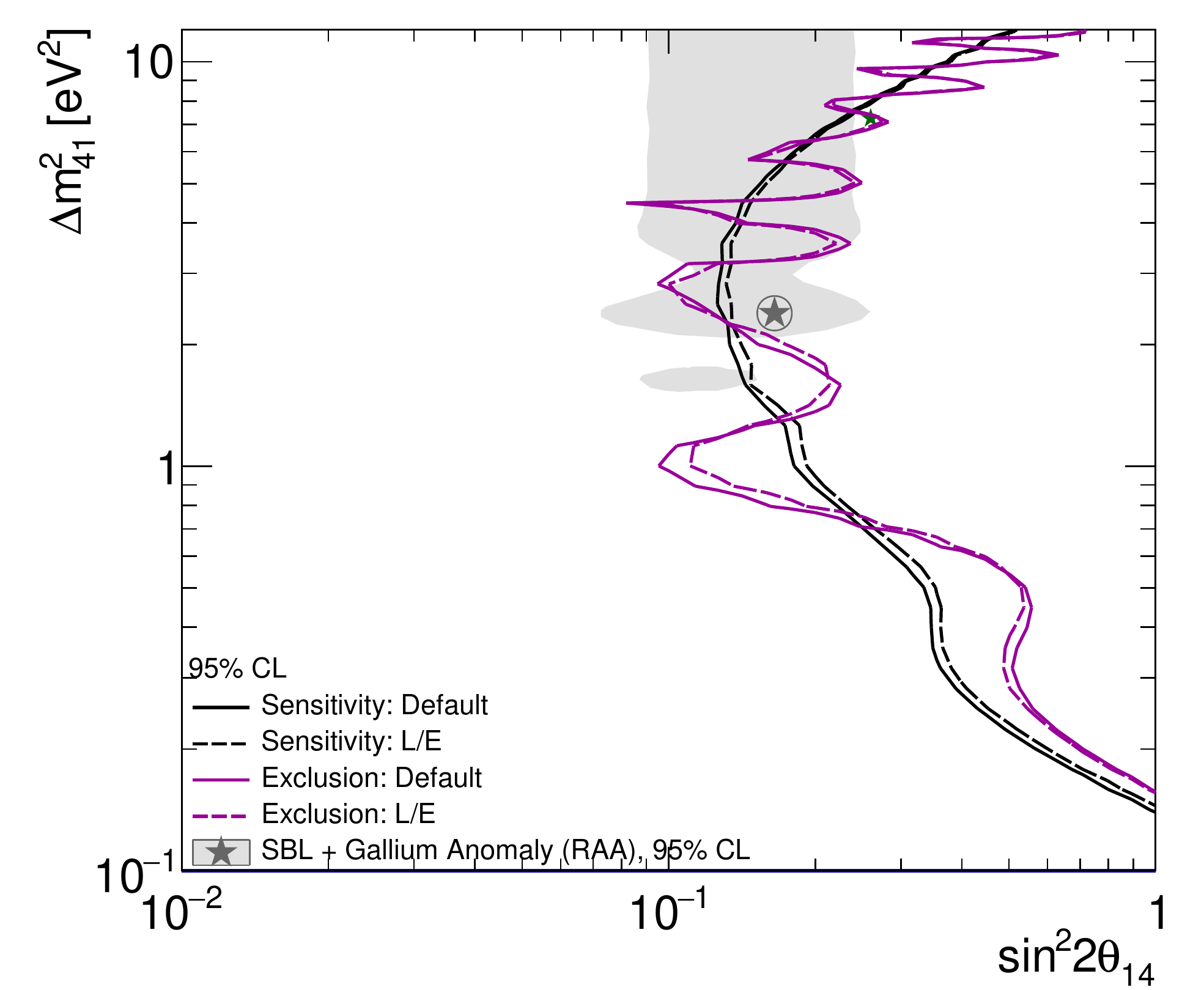}
\caption{A comparison between $L/E$ fit and the default fit performed with PROSPECT first oscillation search data using Gaussian CL$_s$~\cite{Gaussian_CLS} method\footnote{Gaussian CLs method yields similar results to the simulated MC-based method when defining exclusion regions~\cite{Gaussian_CLS,surukuchi2019search}}. Both approaches use a total of 96 bins. Note that this simple test only uses statistical uncertainties and hence shows better coverage than the results shown in Fig.~5 of Ref.~\cite{PROSPECT_First_Search}.}
\label{fig:PROSPECT_LOverE}
\end{figure}

%\section{Oscillation Searches in PROSPECT and STEREO}
It was claimed by the Neutrino-4 experiment on page 27 of Ref.~\cite{Serebrov:2020May} that `... without method of the coherent summation of data by $L/E$ parameter, it is practically impossible to extract the effect of the oscillations from experimental data'.
The coherent summation method as defined in Section 19 of ~\cite{Serebrov:2020May} and reiterated earlier in this paper is a search for oscillations using $L/E$ as a combined baseline-energy parameter. 
Neutrino-4 is the only short baseline reactor neutrino experiment which performed a search for eV-scale sterile neutrinos using this parameter. 
Most other contemporary short baseline experiments including PROSPECT~\cite{PROSPECT_First_Search} and STEREO~\cite{STEREO2019} use a binned test statistic over both baseline and energy without collapsing them into $L/E$ bins. 
Although both methods are expected to yield similar results, the latter method is preferable owing to the fact that systematic effects similar to the ones described in the previous sections can be decoupled between baseline and energy, the two parameters that induce oscillations.
This makes interpreting and implementing of systematic uncertainties transparent and facilitates in avoiding biases that might originate from complicated correlations between baseline and energy.

To demonstrate the similarity of results, we present a comparison between the oscillation fits performed using both $L/E$ parameter and separate $L$ and $E$ parameters.
The latter method is identical to the one used by the PROSPECT experiment in its first search for sterile neutrino oscillations~\cite{PROSPECT_First_Search}. 
We use the dataset which corresponds to 33~(28) live days of reactor-on~(reactor-off) PROSPECT data.
Fig.~\ref{fig:PROSPECT_LOverE} clearly demonstrates that the search for oscillations using $L/E$ as the oscillation parameter yields nearly identical results in both sensitivity and exclusion to the default fit performed by PROSPECT.

\section{Summary}
In conclusion, the sections above outline significant potential issues with the oscillation search performed by the Neutrino-4 collaboration. 
\begin{itemize}
    \item Using STEREO simulated MC datasets, we showed that to unequivocally claim an oscillation signal a proper assignment of significance is required, e.g. through a high-fidelity set of Monte Carlo simulations, and that Wilks' theorem is inadequate.
    \item We have detailed potential oscillation mimicking systematic effects and suggested some instances where more information from the Neutrino-4 collaboration would be beneficial in clarifying the methods used by the collaboration.
    \item Using the PROSPECT dataset as an example, we have also established that the oscillation search using $L$ and $E$ parameters separately is as good as, if not better than, using the $L/E$ parameter, thus calling into question the claim to the contrary made by the Neutrino-4 collaboration.
\end{itemize} 
\bibliographystyle{plain}
\bibliography{references}
\end{document}

%% file: Authors.tex
%\author{PROSPECT and STEREO Collaborations}
%\author{
%PROSPECT$^{p}$ and
%STEREO$^{s}$ collaborations}
%\address{$^{p}$prospect.collaboration@gmail.com}
%\address{$^{s}$http://www.stereo-experiment.org}

\affiliation{Brookhaven National Laboratory, Upton, NY, USA}
\affiliation{Department of Physics, Drexel University, Philadelphia, PA, USA}
\affiliation{George W.\,Woodruff School of Mechanical Engineering, Georgia Institute of Technology, Atlanta, GA USA}
\affiliation{Department of Physics \& Astronomy, University of Hawaii, Honolulu, HI, USA}
\affiliation{Department of Physics, Illinois Institute of Technology, Chicago, IL, USA}
\affiliation{Nuclear and Chemical Sciences Division, Lawrence Livermore National Laboratory, Livermore, CA, USA}
\affiliation{Department of Physics, Le Moyne College, Syracuse, NY, USA}
\affiliation{National Institute of Standards and Technology, Gaithersburg, MD, USA}
\affiliation{High Flux Isotope Reactor, Oak Ridge National Laboratory, Oak Ridge, TN, USA}
\affiliation{Physics Division, Oak Ridge National Laboratory, Oak Ridge, TN, USA}
\affiliation{Department of Physics, Temple University, Philadelphia, PA, USA}
\affiliation{Department of Physics and Astronomy, University of Tennessee, Knoxville, TN, USA}
\affiliation{Institute for Quantum Computing and Department of Physics and Astronomy, University of Waterloo, Waterloo, ON, Canada}
\affiliation{Department of Physics, University of Wisconsin, Madison, Madison, WI, USA}
\affiliation{Wright Laboratory, Department of Physics, Yale University, New Haven, CT, USA}

\author{M.\,Andriamirado}
\affiliation{Department of Physics, Illinois Institute of Technology, Chicago, IL, USA}
\author{A.\,B.\,Balantekin}
\affiliation{Department of Physics, University of Wisconsin, Madison, Madison, WI, USA}
\author{H.\,R.\,Band}
\affiliation{Wright Laboratory, Department of Physics, Yale University, New Haven, CT, USA}
\author{C.\,D.\,Bass}
\affiliation{Department of Physics, Le Moyne College, Syracuse, NY, USA}
\author{D.\,E.\,Bergeron}
\affiliation{National Institute of Standards and Technology, Gaithersburg, MD, USA}
\author{D.\,Berish}
\affiliation{Department of Physics, Temple University, Philadelphia, PA, USA}
\author{N.\,S.\,Bowden}
\affiliation{Nuclear and Chemical Sciences Division, Lawrence Livermore National Laboratory, Livermore, CA, USA}
\author{J.\,P.\,Brodsky}
\affiliation{Nuclear and Chemical Sciences Division, Lawrence Livermore National Laboratory, Livermore, CA, USA}
\author{C.\,D.\,Bryan}
\affiliation{High Flux Isotope Reactor, Oak Ridge National Laboratory, Oak Ridge, TN, USA}
\author{T.\,Classen}
\affiliation{Nuclear and Chemical Sciences Division, Lawrence Livermore National Laboratory, Livermore, CA, USA}
\author{A.\,J.\,Conant}
\affiliation{George W.\,Woodruff School of Mechanical Engineering, Georgia Institute of Technology, Atlanta, GA USA}
\author{G.\,Deichert}
\affiliation{High Flux Isotope Reactor, Oak Ridge National Laboratory, Oak Ridge, TN, USA}
\author{M.\,V.\,Diwan}
\affiliation{Brookhaven National Laboratory, Upton, NY, USA}
\author{M.\,J.\,Dolinski}\affiliation{Department of Physics, Drexel University, Philadelphia, PA, USA}
\author{A.\,Erickson}
\affiliation{George W.\,Woodruff School of Mechanical Engineering, Georgia Institute of Technology, Atlanta, GA USA}
\author{B.\,T.\,Foust}
\affiliation{Wright Laboratory, Department of Physics, Yale University, New Haven, CT, USA}
\author{J.\,K.\,Gaison}
\affiliation{Wright Laboratory, Department of Physics, Yale University, New Haven, CT, USA}
\author{A.\,Galindo-Uribarri}\affiliation{Physics Division, Oak Ridge National Laboratory, Oak Ridge, TN, USA} \affiliation{Department of Physics and Astronomy, University of Tennessee, Knoxville, TN, USA}
\author{C.\,E.\,Gilbert}\affiliation{Physics Division, Oak Ridge National Laboratory, Oak Ridge, TN, USA} \affiliation{Department of Physics and Astronomy, University of Tennessee, Knoxville, TN, USA}
\author{B.\,T.\,Hackett}\affiliation{Physics Division, Oak Ridge National Laboratory, Oak Ridge, TN, USA} \affiliation{Department of Physics and Astronomy, University of Tennessee, Knoxville, TN, USA}
\author{S.\,Hans}\affiliation{Brookhaven National Laboratory, Upton, NY, USA}
\author{A.\,B.\,Hansell}
\affiliation{Department of Physics, Temple University, Philadelphia, PA, USA}
\author{K.\,M.\,Heeger}
\affiliation{Wright Laboratory, Department of Physics, Yale University, New Haven, CT, USA}
%\author{B.\,Heffron}\affiliation{Physics Division, Oak Ridge National Laboratory, Oak Ridge, TN, USA} \affiliation{Department of Physics and Astronomy, University of Tennessee, Knoxville, TN, USA}
\author{D.\,E.\,Jaffe}
\affiliation{Brookhaven National Laboratory, Upton, NY, USA}
\author{X.\,Ji}
\affiliation{Brookhaven National Laboratory, Upton, NY, USA}
\author{D.\,C.\,Jones}
\affiliation{Department of Physics, Temple University, Philadelphia, PA, USA}
\author{O.\,Kyzylova}\affiliation{Department of Physics, Drexel University, Philadelphia, PA, USA}
\author{C.\,E.\,Lane}\affiliation{Department of Physics, Drexel University, Philadelphia, PA, USA}
\author{T.\,J.\,Langford}
\affiliation{Wright Laboratory, Department of Physics, Yale University, New Haven, CT, USA}
\author{J.\,LaRosa}
\affiliation{National Institute of Standards and Technology, Gaithersburg, MD, USA}
\author{B.\,R.\,Littlejohn}
\affiliation{Department of Physics, Illinois Institute of Technology, Chicago, IL, USA}
\author{X.\,Lu}\affiliation{Physics Division, Oak Ridge National Laboratory, Oak Ridge, TN, USA} \affiliation{Department of Physics and Astronomy, University of Tennessee, Knoxville, TN, USA}
\author{J.\,Maricic}\affiliation{Department of Physics \& Astronomy, University of Hawaii, Honolulu, HI, USA}
\author{M.\,P.\,Mendenhall}\affiliation{Nuclear and Chemical Sciences Division, Lawrence Livermore National Laboratory, Livermore, CA, USA}
\author{A.\,M.\,Meyer}\affiliation{Department of Physics \& Astronomy, University of Hawaii, Honolulu, HI, USA}
\author{R.\,Milincic}\affiliation{Department of Physics \& Astronomy, University of Hawaii, Honolulu, HI, USA}
\author{I.\,Mitchell}\affiliation{Department of Physics \& Astronomy, University of Hawaii, Honolulu, HI, USA}
\author{P.\,E.\,Mueller}\affiliation{Physics Division, Oak Ridge National Laboratory, Oak Ridge, TN, USA} 
\author{H.\,P.\,Mumm}
\affiliation{National Institute of Standards and Technology, Gaithersburg, MD, USA}
\author{J.\,Napolitano}
\affiliation{Department of Physics, Temple University, Philadelphia, PA, USA}
\author{C.\,Nave}\affiliation{Department of Physics, Drexel University, Philadelphia, PA, USA}
\author{R.\,Neilson}\affiliation{Department of Physics, Drexel University, Philadelphia, PA, USA}
\author{J.\,A.\,Nikkel}
\affiliation{Wright Laboratory, Department of Physics, Yale University, New Haven, CT, USA}
\author{D.\,Norcini}
\affiliation{Wright Laboratory, Department of Physics, Yale University, New Haven, CT, USA}
\author{S.\,Nour}
\affiliation{National Institute of Standards and Technology, Gaithersburg, MD, USA}
\author{J.\,L.\,Palomino-Gallo}
\affiliation{Department of Physics, Illinois Institute of Technology, Chicago, IL, USA}
\author{D.\,A.\,Pushin}\affiliation{Institute for Quantum Computing and Department of Physics and Astronomy, University of Waterloo, Waterloo, ON, Canada}
\author{X.\,Qian}
\affiliation{Brookhaven National Laboratory, Upton, NY, USA}
\author{E.\,Romero-Romero}\affiliation{Physics Division, Oak Ridge National Laboratory, Oak Ridge, TN, USA} \affiliation{Department of Physics and Astronomy, University of Tennessee, Knoxville, TN, USA}
\author{R.\,Rosero}
\affiliation{Brookhaven National Laboratory, Upton, NY, USA}
\author{P.\,T.\,Surukuchi}
\affiliation{Wright Laboratory, Department of Physics, Yale University, New Haven, CT, USA}
\author{M.\,A.\,Tyra}
\affiliation{National Institute of Standards and Technology, Gaithersburg, MD, USA}
\author{R.\,L.\,Varner}\affiliation{Physics Division, Oak Ridge National Laboratory, Oak Ridge, TN, USA} 
\author{D.\,Venegas-Vargas}\affiliation{Physics Division, Oak Ridge National Laboratory, Oak Ridge, TN, USA} \affiliation{Department of Physics and Astronomy, University of Tennessee, Knoxville, TN, USA}
\author{P.\,B.\,Weatherly}\affiliation{Department of Physics, Drexel University, Philadelphia, PA, USA}
\author{C.\,White}
\affiliation{Department of Physics, Illinois Institute of Technology, Chicago, IL, USA}
\author{J.\,Wilhelmi}
\affiliation{Wright Laboratory, Department of Physics, Yale University, New Haven, CT, USA}
\author{A.\,Woolverton}\affiliation{Institute for Quantum Computing and Department of Physics and Astronomy, University of Waterloo, Waterloo, ON, Canada}
\author{M.\,Yeh}
\affiliation{Brookhaven National Laboratory, Upton, NY, USA}
\author{A.\,Zhang}
\affiliation{Brookhaven National Laboratory, Upton, NY, USA}
\author{C.\,Zhang}
\affiliation{Brookhaven National Laboratory, Upton, NY, USA}
\author{X.\,Zhang}
\affiliation{Nuclear and Chemical Sciences Division, Lawrence Livermore National Laboratory, Livermore, CA, USA}
\collaboration{PROSPECT collaboration}
\email{prospect.collaboration@gmail.com}

\newcommand{\STEREO}{\textsc{Stereo} }

\newcommand{\MPIK}{\affiliation{Max-Planck-Institut f\"ur Kernphysik, Saupfercheckweg 1, 69117 Heidelberg, Germany}}
\newcommand{\LAPP}{\affiliation{Univ.~Grenoble Alpes, Universit\'e Savoie Mont Blanc, CNRS/IN2P3, LAPP, 74000 Annecy, France}}
\newcommand{\LPSC}{\affiliation{Univ.~Grenoble Alpes, CNRS, Grenoble INP, LPSC-IN2P3, 38000 Grenoble, France}}
\newcommand{\CEA}{\affiliation{IRFU, CEA, Universit\'e Paris-Saclay, 91191 Gif-sur-Yvette, France}}
\newcommand{\ILL}{\affiliation{Institut Laue-Langevin, CS 20156, 38042 Grenoble Cedex 9, France}}

\author{H.~Almaz\'an}\MPIK
\author{A.~Bonhomme}\MPIK\CEA
\author{C.~Buck}\MPIK
\author{P.~del~Amo~Sanchez}\LAPP
\author{I.~El~Atmani}\altaffiliation[Present address: ]{Hassan II University, Faculty of Sciences, A\"in Chock, BP 5366 Maarif, Casablanca 20100, Morocco}\CEA
\author{L.~Labit}\LAPP
\author{J.~Lamblin}\LPSC
\author{A.~Letourneau}\CEA
\author{D.~Lhuillier}\CEA
\author{M.~Licciardi}\LPSC
\author{T.~Materna}\CEA
\author{H.~Pessard}\LAPP
\author{J.-S.~R\'eal}\LPSC
\author{C.~Roca}\MPIK
\author{R.~Rogly}\CEA
\author{V.~Savu}\CEA
\author{S.~Schoppmann}\MPIK 
\author{T.~Soldner}\ILL
\author{A.~Stutz}\LPSC
\author{M.~Vialat}\ILL

\collaboration{\STEREO Collaboration}
\homepage{http://www.stereo-experiment.org}